\documentclass[twocolumn,superscriptaddress,pra,aps,showpacs]{revtex4}
\usepackage{amssymb}

\usepackage{amsfonts}
\usepackage{amsmath}
\usepackage{epsfig}

\begin{document}

\title{Classical Trajectory Diagnosis of Finger-Like Pattern in
the Correlated Electron Momentum Distribution for Helium Double
Ionization}
\author{D.F. Ye}
\affiliation{ Center for Applied Physics and Technology, Peking
University, 100084, Beijing, China} \affiliation{Institute of
Applied Physics and Computational Mathematics, P.O.Box 100088,
Beijing, China} \affiliation{Graduate School, China Academy of
Engineering Physics, Beijing 100088, China}
\author{X. Liu}
\affiliation{State Key Laboratory of Magnetic Resonance and Atomic
and Molecular Physics, Wuhan Institute of Physics and Mathematics,
Chinese Academy of Sciences, Wuhan 430071, China}
\author{J. Liu$^{1,2}$}
\begin{abstract}
With a semiclassical quasistatic model we identify the distinct
roles of nuclear Coulomb attraction, final state electron repulsion
and electron-field interaction in forming the finger-like (or
V-shaped) pattern in the correlated electron momentum distribution
for Helium double ionization [Phys. Rev. Lett. \textbf{99}, 263002;
\emph{ibid}, 263003 (2007)]. The underlying microscopic trajectory
configurations responsible for asymmetric electron energy sharing
after electron-electron collision have been uncovered and
corresponding sub-cycle dynamics are analyzed. The correlation
pattern is found to be sensitive to the transverse momentum of
correlated electrons.
\end{abstract}

 \maketitle

Nonsequential double ionization (NSDI) of atoms subject to
ultrashort intense laser pulses attracts constant interests because
it is a prototype model for the study of three-body Coulomb problem
intervened by the highly unperturbed interaction of the electrons
with the strong laser field. Until recently it has been consensus
that rescattering is the dominant mechanism for NSDI \cite{focus}.
In this three-step mechanism, the first electron is freed by a
quasi-static tunneling ionization, and is driven back to its parent
ion and imparts part of energy to dislodge a second electron.

The electron recollision picture as a  cornerstone  of the
rescattering mechanism inspires the further investigations that
achieve insight into the microscopic dynamics of the ionization
process on the timescale of subfemtosecond. The advent of
experimental techniques, represented by the sophisticated Cold
Target Recoil Ion Momentum Spectroscopy (COLTRIMS), combined with
high-repetition-rate lasers, has to a large extent facilitated this
type of study. For example, the observed double hump structure in
the
ion momenta \cite{doublehump} and the electron momenta correlation \cite%
{eecorrelation} parallel to the field gave solid evidence of the time delay
introduced by the rescattering process and the emission time of both
electrons close to the zero crossing of the oscillating field.

Despite the great success of this picture, a comprehensive
understanding of the microscopic dynamics in this recollision
process is far from being complete. Indeed, new high resolution and
high statistics COLTRIMS experiments on double ionization of helium
are performed independently by two groups and a striking finger-like
(or V-shaped) structure is observed \cite%
{fingerexp1,fingerexp2} in the correlated electron momenta parallel
to the laser polarization, in qualitative  accordance with the
prediction of S-matrix approach \cite{figueira1,figueira2,figueira3}
and quantum mechanical calculation \cite{fingertheory}. These high
resolution and high statistics COLTRIMS experiments of double
ionization (DI) provide benchmark data  for comprehensive
theoretical treatments.

In this letter, by exploiting  an \textit{%
ab initio} 3D semiclassical model, we have reproduced essentially
all the experimental characteristics, and identified the distinct
roles of nuclear Coulomb attraction, final state electron repulsion
and electron-field interaction in the formation of the  finger-like
structure. Furthermore, classical trajectory (CT) analysis
facilitates to unveil the  sub-cycle microscopic dynamics behind the
finger-like structure.

Compared with other approximate approaches extensively employed in strong
field double ionization, e.g. the one dimensional quantum model \cite%
{1quantum}, strong field S-matrix calculation
\cite{figueira1,figueira2,figueira3,wbecker} and simplified
classical methods \cite{eberly}, our semiclassical model has the
advantage that all the effects determining the DI ionization
process, such as the quantum tunneling, the effective interactions
between particles and the laser field as well as the Coulomb
focusing effect, can be fully included, while keeping the
computational capacity still accessible. The model has achieved
great success in explaining various DI phenomena \cite{fu1},
including the excessive DI yield, the recoil momentum distribution
of doubly ionized ions, momentum correlation between two emitted
electrons, and the energy spectra and angular distribution of
photoelectrons.

For simplicity, we just briefly present the theoretical methodology
here. We consider a helium atom interacting with an infrared laser
pulse. When the laser field is strong enough, one electron is
released at the outer edge of the suppressed Coulomb potential
through quantum tunneling with a rate given by the ADK formula
\cite{adk}. The tunneled electron has a Gaussian-like  distribution
on transverse velocity and zero longitudinal velocity \cite{fu1}.
For the bound electron, the initial position and momentum are
depicted by single-electron microcanonical distribution (SMD)
\cite{smd}. The subsequent evolution of the two electrons with the
above initial
conditions are governed by Newton's equations of motion: $\frac{d^{2}\mathbf{%
r}_{i}}{dt^{2}}=\mathbf{\epsilon }(t)-\bigtriangledown
_{r_{i}}(V_{ne}^{i}+V_{ee}).$ Here the index i denotes the two
different
electrons, $V_{ne}^{i}=-\frac{2}{%
\left\vert \mathbf{r}_{i}\right\vert }$ and $V_{ee}=\frac{1}{\left\vert \mathbf{r}%
_{1}-\mathbf{r}_{2}\right\vert }$, are Coulomb interaction between
nucleus and electrons and between two electrons, respectively.

\begin{figure}[t]
\begin{center}
\rotatebox{0}{\resizebox *{5.0cm}{6.0cm} {\includegraphics [bb= 430
17 615 240] {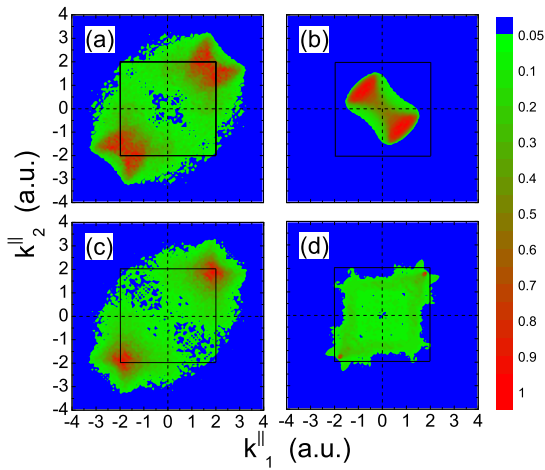}}}
\end{center}
\caption{(Color online). (a) Distribution of correlated electron
momenta along the laser polarization for Helium DI irradiated by
800nm, $4.5\times 10^{14}$W/cm$^{2}$ laser pulses. The black box
indicates the classical limit of the tunneled electron momentum. The
most important experimental
feature, i.e., the finger-like structure beyond the limit of $2\protect\sqrt{%
U_{p}}$ is well reproduced by our model calculation. The model
calculations under various circumstances yield very different
momentum distribution patterns(see text for details): (b) the laser
field is removed and the tunneled electrons are replaced by a beam
of projectile electrons; (c) the electron-electron Coulombic
interaction is replaced with a Yukawa potential; (d) the nuclear
Coulomb potential is softened.} \label{finger}
\end{figure}

The above Newtonian equations are solved by employing the standard
4-5th Runge-Kutta algorithm and DI events are identified by energy
criterion. In our calculations, more than $10^{7}$ weighted (i.e.,
by the tunneling rate ) classical two-electron trajectories are
traced until one electron moves to such a position that
$r_{i}>200a.u.$. This results in more than 10$^4$ DI events for
statistics.

The resulting electron momentum distribution, calculated with this
semiclassical model for the same parameters as in the experiment \cite%
{fingerexp1}, is shown in Fig. \ref{finger}(a). The calculation
reproduces many key features observed in the experiment, including
the emission of the two electrons primarily into the same
hemisphere,  the small circular accumulation around the zero
momentum surrounded by four elliptical hard-to-reach regime, and
more importantly, the finger-like
structure beyond the limit of $2\sqrt{U_{p}}$.  Here $%
U_{p}=\epsilon_0 ^{2}/(4\omega ^{2})$, the ponderomotive energy,
refers to the cycle averaged quiver energy of a free electron in an
oscillating electric field.

The off-diagonal and beyond-limit properties of the finger-like
structure is striking and contradicts to the traditional scenario on
DI, in which the electron momentum at the time of emission is
assumed to be small and the postcollision electron-electron
interaction supposed to be weak. Therefore, the parallel momentum
$k^{||}_{1,2}$ of each electron results exclusively from the
acceleration in the optical field: $
k^{||}_{1,2}=2\sqrt{U_p}\sin\omega t_{ion}$ with ionization time
$t_{ion}$\cite{new1}. Within this scenario $2\sqrt{U_{p}}$ should be
the maximal momentum and the momentum distribution favors
accumulation in the diagonal zone  because the electrons are emitted
nearly simultaneously.

We now proceed to explore the physical effects that give rise to
this peculiar finger-like structure. In the context of strong field
double ionization, there are essentially three major effects that
may play significant role in the double electron emission dynamics:
electron-laser field interaction that occurs throughout the DI
process, electron-nuclear Coulomb interaction in the post-collision
duration and the inter-electron Coulomb repulsion which becomes
significant when both electrons get close. Below we investigate all
three interactions and clarify their distinct roles in the formation
of the finger-like structure.

The first step is to check the role of the external laser field and
an additional calculation is thus performed, in which the laser
field is intentionally removed and the tunneled electrons are
replaced by a beam of projectile ones with incident energy of
$3.17U_{p}$, corresponding to the maximal kinetic energy of the
tunneled electrons upon recollision. The result is shown in Fig.
\ref{finger}(b). Two significant differences from the complete model
calculation in Fig. \ref{finger}(a) are found: (i) the finger-like
structure beyond the limit of $2\sqrt{U_{p}}$ completely disappears;
(ii) the two emitted electrons tend to distribute in the second and
fourth quadrants of the parallel momentum plane, indicating that the
incident electron transfers much of its momentum to the bound one
while itself is back-scattered into the opposite direction. The
comparison between Fig. \ref{finger}(a) and (b) shows the most
important role of the laser field in turning the two back-to-back
emitted electrons into the same direction and accordingly the
finger-like structure.

The next step concerns the question if this finger-like structure is
a fingerprint of a strong inter-electron correlation among the
ionizing electrons. We have performed another calculation in which
the final-state electron Coulomb repulsion has been deliberately
neglected by
replacing the electron Coulombic interaction $V_{ee}=\frac{1}{%
\left\vert \mathbf{r}_{1,2}\right\vert }$ with Yukawa repulsion potential of
the form $V_{\mathrm{ee}}=\exp [-\lambda r_{b}]/r_{b}$, where $r_{b}=%
\sqrt{ \left\vert \mathbf{r}_{1,2}\right\vert ^{2}+b^{2}}$, $\lambda
=5.0$ and $b=0.2$. The result of this calculation [Fig.
{\ref{finger}}(c)] shows that the prominent finger-like structure is
to a large extent reduced and thus provides a clear evidence that
the final-state electron correlation plays a significant role.

Last but not the least procedure is to justify the role of the
electron-nuclear interaction, which is commonly believed to be the
main reason for the recoil collision in field-free (e, 2e) process.
This interaction was also suggested to be the very ingredient for
the field-assisted recoil collision in the context of intense field
DI of atomic helium \cite{fingerexp1}. Accordingly,
an additional calculation, in which we soften the nuclear Coulomb attraction by employing $%
V_{ne}^{i}=-2/\sqrt{{%
\left\vert \mathbf{r}_{i}\right\vert }^{2}+a^{2}}$, where $a$ is
chosen as 1.0 to match the ground state energy of He$^{+}$, is
performed.

Physically, the shielding of nuclear potential would to a great extent
diminish the Coulomb focusing effect that have significant effects upon both
electrons. Clearly, a Coulombic potential would attract the tunneled
electron more dramatically when it moves near the atomic core. Such strong
attraction may unambiguously bring the tunneled electron to share more
kinetic energy with the bound one. For the bound electron, after achieving
considerable transferred momentum upon collision, it may elastically
backscatter from the Coulombic core on its way out of the atom. This double
scattering process is coined as recoil collision \cite{recoil} and was
routinely found in traditional electron impact ionization experiments,
especially when the projectile electron possesses the energy of only a few
times of the binding energy of the inner one. The result shown in Fig. {\ref%
{finger}}(d) indicates that Coulomb focusing effect is decisive for the
production of the electrons with high energy, and thus for the finger-like
structure.

\begin{figure}[t]
\begin{center}
\rotatebox{0}{\resizebox *{7.0cm}{5.5cm} {\includegraphics
{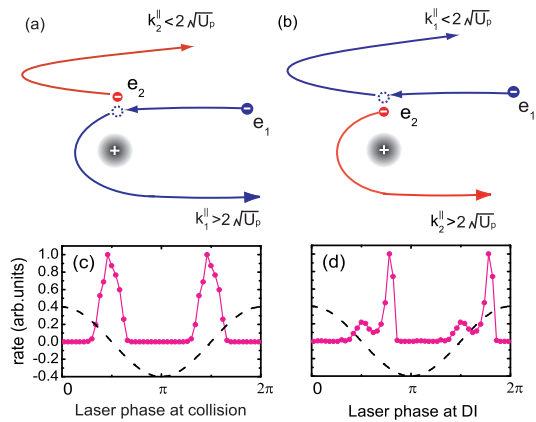}}}
\end{center}
\caption{(Color online). {(a-b) Two trajectory configurations
responsible for the finger like structure. DI yield versus laser
phase at recollision (c) and at DI moment(d). Here, the statistics
is only collected for the  DI trajectories that contribute to the
prominent finger-like structure, i.e., the regimes of $1.5$ $a.u.<|
k_{i}^{||} | <2.0$ $a.u.$ and $2.5 $ $a.u.<| k_{j}^{||}| <3.0$
$a.u.$, where $i,j=1,2, i\neq j$. The dashed curves represent the
laser field for guiding eyes.}} \label{scheme}
\end{figure}

The dynamics behind the finger-like structure observed in the
experiments was supposed to be related with the recoil collision in
the presence of the external laser field \cite{fingerexp1}. This
conjecture has been justified by a 2-body 3D quantum mechanical
simulation of reduced dimensionality \cite{Ruiz2006}. However the
dynamical details of such collision is hardly explored in the
quantum mechanical treatment. In the remaining part of this letter,
we proceed to unveil the collision dynamics of the DI electrons,
especially of those that contribute to the finger-like structure
beyond the limit $2\sqrt{U_{p}}$. With the CT approach, this becomes
possible by back-analyzing the history of the DI events of interest.

It has been recognized that the characteristics of the DI trajectory
can be well represented by the recollision and DI time
\cite{eberly}. We thus provide such an information for the
trajectories that contribute to the finger-like structure in Fig.
\ref{scheme}(c) and (d). It is found that, in Fig. \ref{scheme}(c),
the electron pairs contributing to the finger-like structure tend to
encounter right at zero field. Within rescattering picture, this can
be understood as that these trajectories include the most energetic
collisions for which the tunneled electrons are released at the
laser phase of about $17^{\circ }$, and return around the zero
crossing of the electric field at $270^{\circ }$ with maximal energy
of 3.17$U_{p}$ \cite{new1,paulus}.

Upon recollision, the bound electron may be directly freed, a
process termed as collision ionization (CI),  or be excited and
subsequently ionized by the next field maximum, known as
collision-excitation ionization (CEI) \cite{CEI}. The smaller and
larger peaks in Fig. \ref{scheme}(d) correspond to these two
mechanisms, respectively. The smaller ones around zero field
represents the electron pairs emitted after a very short
thermalization process ($\sim attosecond$) \cite{therm}. While the
larger peaks correspond to CEI events that are ionized after a few
optical cycles delay leading to a $0.3\pi$ phase difference from the
collision phase peaks \cite{prl}. The above calculation indicates
that both CI and CEI types of DI events contribute to the
finger-like structure. The time delay in CEI could lead to an
off-diagonal momentum distribution apart from $k_1^{\parallel} =
k_2^{\parallel}$, but, nevertheless, cannot account for the excess
momentum that spills over the limit of $2\sqrt{U_{p}}$.

To further unveil the microscopic mechanism underlying this unusual
pattern, we collect all trajectories that constitute the finger-like
structure and  trace back their dynamical evolution at collision and
post-collision. With the CT diagnosis we sketch two types of
trajectory configurations that are responsible for the asymmetric
electron energy sharing after electron-electron collision and leads
to the finger-like structure beyond the limit of $2\sqrt{U_p}$ (see
Fig. \ref{scheme}(a) and (b), correspond to type-I and II
configurations, respectively). While the tunneled electron is driven
back to the nucleus by the laser field, the field strength reduces
to zero and the collision is essentially a field free three-body
system under the pure Coulomb potential. For type-I trajectories, as
shown in Fig. \ref{scheme}(a), the electron-electron collision near
nucleus could lead to following consequences: the second electron
acquires considerable momentum from the returned electron and emits
in the forward direction, while the returned electron is slowed
down. Under the influence of the nuclear attraction, the latter is
transferred to a hyperbolic orbit around the nucleus with a large
scattering angle. In this way, the returned electron reverses its
direction in a time scale of attoseconds after the collision. Notice
that meanwhile the laser field changes its direction. As the
returned electron has nonzero residual momentum parallel to the
instantaneous field direction and is further accelerated by the
field, its final longitudinal momentum is expected to be above the
limit $2\sqrt{U_p}$. For the second electron, after collision its
initial momentum is opposite to the instantaneous field direction,
one expects that its final longitudinal momentum is below the limit
$2\sqrt{U_p}$. In Fig. \ref{scheme}(b) where a type-II trajectory is
schematically shown, the situation is similar except that the roles
of the two electrons have exchanged: Under assistance of nuclear
Coulomb attraction, the second electron acquires a nonzero momentum
parallel to the instantaneous field and therefore emits with a
longitudinal momentum larger than $2\sqrt{U_{p}}$. Although the
returned electron is slowed down by electron-electron repulsion, it
still has a residual momentum opposite to the instantaneous field,
resulting in a final longitudinal momentum below $2\sqrt{U_p}$. Our
statistics reveals that the DI events in the finger-like structure
consists of 30$\%$ type-I configuration and 70$\%$ type-II
configuration.

\begin{figure}[t]
\begin{center}
\rotatebox{0}{\resizebox *{5.0cm}{6.0cm} {\includegraphics [bb= 425
30 615 245] {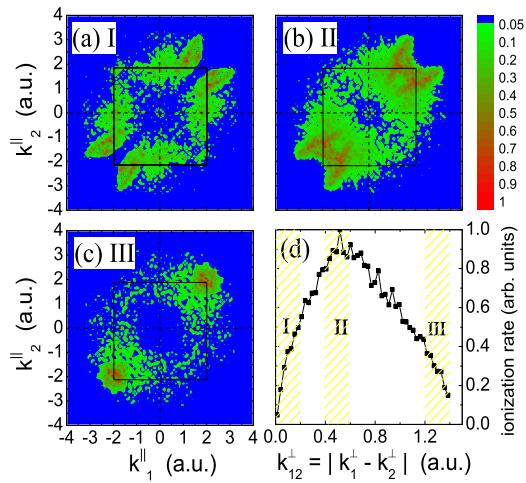}}}
\end{center}
\caption{(Color online). Correlated parallel momentum distributions
with additional conditions on the relative perpendicular momentum
between two electrons, i.e., (a) 0 a.u. $\leqslant k_{12}^{\bot
}$\textbf{\ }$\leqslant$
0.2 a.u., (b) 0.4 a.u. $\leqslant k_{12}^{\bot }$\textbf{\ }$%
\leqslant$ 0.6 a.u., (c) 1.2 a.u. $\leqslant k_{12}^{\bot }$\textbf{\ }%
$\leqslant$ 1.4 a.u.. (d) The overall relative perpendicular
momentum distribution.} \label{perp}
\end{figure}

The above analysis reveals  that the electron-electron collision
assisted by the nuclear attraction is crucial  for the emergence of
finger-like structure. In certain cases that the collision is
strong, the finger-like structure should be more prominent. To test
this idea, we impose  an additional confinement on our statistics to
observe the variation of the correlated momentum patterns with
respect to the relative perpendicular momentum between two
electrons. The results are presented in Fig. \ref{perp}.

It is shown in Fig. \ref{perp}(d) that the relative perpendicular
momentum $k_{12}^{\bot }$ for all DI events distributes over an
interval of $[0,1.4]$ and exhibits a notable accumulation around 0.5
a.u.. Moreover, the finger-like pattern is more prominent for the
case of small relative perpendicular momentum (see Fig.
\ref{perp}(a)). As we increase the value of $k_{12}^{\bot }$ the two
finger patterns start to merge (Fig. \ref{perp}(b)) and finally
totally disappear (Fig. \ref{perp}(c)). The above result is in
agreement with the experimental observations
\cite{fingerexp2,staudte}, and provides a good explanation for the
1D quantum calculation in \cite{1quantum}. Due to dimensional
restriction, the quantum calculation made there greatly magnifies
the electron-electron collision effects. As a result, a
butterfly-like structure similar to Fig. \ref{perp}(a) emerges (see
Fig. 1 of \cite{1quantum} ).

In summary, with a semiclassical quasistatic model we have made the
CT diagnosis on the finger-like structure observed in recent
experiments, and unveiled the microscopic mechanism behind the
striking pattern. Our results suggest that the finger-like structure
is generated by the interplay of backscattering of the returning
electron  and the electron-electron Coulomb repulsion.

We are grateful to Dr. A. Staudte for making their experimental
results accessible prior to publication and many useful discussions.
This work is supported by NNSF of China (No. 10725521 and 10674153),
the National Fundamental Research Programme of China No.
2006CB806000, 2006CB921400, 2007CB814800 and CAEP Foundation No.
2006Z0202.

\end{document}